\documentstyle[aps,multicol,epsf,rotate,prl,psfig]{revtex}
\begin{document}
\title{Characteristic length scale and energy of a vortex line in a dilute, 
superfluid Fermi gas}
\author{\O. Elgar\o y}  
\address{Institute of Astronomy, University of Cambridge, Madingley Road, 
Cambridge CB3 0HA, UK 
}
\maketitle

\begin{abstract}

We calculate the characteristic length scale of  
a vortex line in a dilute, superfluid gas of fermionic atoms, and 
find that it is in  general smaller, and has a weaker density 
dependence than the BCS coherence length.  Taking this into account, we find the 
energy of a vortex line to be 
larger than has previously been estimated.  As a consequence, the critical frequency for 
formation of vortices in trapped 
fermion gases can be significantly larger than earlier calculations have suggested.  
\end{abstract}

\begin{multicols}{2}

The transition to a superfluid Bardeen-Cooper-Schrieffer (BCS) state at 
low temperatures is a generic feature of  Fermi 
systems with an attractive effective interaction.  The low-temperature 
superconductors and liquid $^3{\rm He}$ are the most familiar examples, 
but the phenomenon can also be seen in more exotic systems like 
atomic nuclei, and superfluids are also thought to be important in neutron 
stars \cite{mhj}.    
Shortly after Bose-Einstein 
condensation in a dilute gas of $^{87}{\rm Rb}$ atoms was achieved 
\cite{becfirst}, the possibility of observing the BCS transition 
to a superfluid state in a dilute gas of trapped fermionic atoms was 
proposed \cite{bcsfirst}.  Trapping and cooling of fermionic 
alkali atoms have now been achieved, reaching temperatures as 
low as $\sim T_F/4$ for $^{40}{\rm K}$ \cite{kalium1,kalium2} and $^6{\rm Li}$ 
\cite{litium1,litium2,litium3}, where $T_F$ is the Fermi temperature.  
The typical  transition 
temperature for weakly interacting fermions is $T_c \sim 10^{-3}T_F$, 
but one hopes to create systems with $T_c \sim 10^{-1}T_F$ in the 
experiments.  If temperatures below $T_c$ can be reached, a new and 
exciting laboratory for studying properties of fermion superfluids will be at our 
disposal.  Finding experimental signatures of the BCS transition in these 
systems is therefore important, and considerable effort has been invested in this problem 
\cite{bcswork}.  

The study of quantized vortices in Bose-Einstein condensates 
has led to many interesting results \cite{vortexbec}, 
and recently the vortex state in trapped fermionic gases has  
also been studied \cite{rodriguez,bruun} as a means for detecting 
a superfluid state in these systems.  A fundamental quantity in these 
considerations is the critical rotational frequency for the creation of 
a vortex line with one quantum of circulation.  The gases in the 
experiments are inhomogeneous because of the trapping potential, but 
if the vortex is produced near the center of the trap, one can estimate the 
critical frequency  starting from the energy per unit length, ${\cal E}_v$, 
of a vortex in homogeneous matter \cite{bruun,lundh}.
Recently, Bruun and Viverit \cite{bruun} calculated ${\cal E}_v$ 
at zero temperature, assuming that the size of the vortex core, the region 
where the order parameter drops from its value in homogeneous matter to zero, 
is given by the BCS coherence length $\xi_0$.  However, several studies of 
vortices in low-temperature superconductors \cite{kramerpesch,gygi2,gygi} have 
found that the formation of bound quasiparticle states localized near the center 
of the vortex  can cause the size of the vortex core to be much smaller than  $\xi_0$
at low temperatures.  The same effect has been found for vortices in 
superfluid neutron matter \cite{deblasio1,deblasio2}, and we will in this 
Letter argue that it is also present in dilute gases, and that this 
can lead to a significantly larger vortex energy than that found in \cite{bruun}.

We consider a dilute, homogeneous gas of fermionic  atoms in two different 
hyperfine states, held at temperature 
$T=0$.  In this limit the Pauli principle dictates that the effective 
$s$-wave interaction between two atoms in the same hyperfine state vanish.  
The interaction between two atoms in different states can be approximated 
by the $s$-wave scattering length $a$.  For a negative scattering length, 
the interaction is attractive and if the number of particles in the 
two states is the same, the $T=0$ ground state of the gas is superfluid.  
Theoretically, the formation of a superfluid state is signaled by a 
non-zero value for the energy gap $\Delta_0$, which is proportional to 
the critical temperature $T_c$.  In the dilute gas limit $k_F|a| \ll 1$, 
where $k_F$ is the Fermi wavenumber, it is given by 
\begin{equation} 
\Delta_0 = \left(\frac{2}{e}\right)^{7/3}\epsilon_F e^{-\pi/2k_F|a|},
\label{eq:wcgap}
\end{equation}
where $\epsilon_F=\hbar^2k_F^2/2m_a$, $m_a$ being the mass of an atom, 
is the Fermi energy \cite{gorkov,heiselberg}.  
Corrections from the so-called induced interaction modify the numerical 
prefactor in this result, but the dependence on $k_F$ and $|a|$ 
remains the same \cite{gorkov,heiselberg}.

The energy per unit length of a vortex line with respect to the homogeneous 
ground state of the superfluid can be estimated by 
adding the kinetic energy associated with the velocity field around 
the vortex, and the loss of condensation energy in the vortex core \cite{bruun}.  
In this approximation the vortex core is treated as a cylindrical 
column of normal matter with radius equal to the BCS coherence length 
$\xi_0 =\hbar v_F/\pi\Delta_0 $, where $v_F = \hbar k_F/m_a$.  
The result is 
\begin{equation}
{\cal E}_v = \frac{\pi\hbar^2 n_\sigma}{2m_a}
\ln\left(D\frac{R_c}{\xi_0}\right),  
\label{eq:simplee}
\end{equation}
where $n_\sigma = k_F^3/6\pi^2$ is the number density of one hyperfine 
state, and $D=1.36$.  
The energy can also be calculated from Ginzburg-Landau theory, 
and the result turns out to be of the same form, but with $D=1.65$ \cite{bruun}, 
again assuming that $\xi_0$ is the characteristic length scale for the vortex.    
However, at low temperatures the microscopic properties of the system are important, 
and, as we will argue below, the existence of quasiparticle states localized 
near the center of the vortex has to be taken into account at very low temperatures.    

\begin{figure} 
\begin{center}
{\centering
\mbox
{\psfig{figure=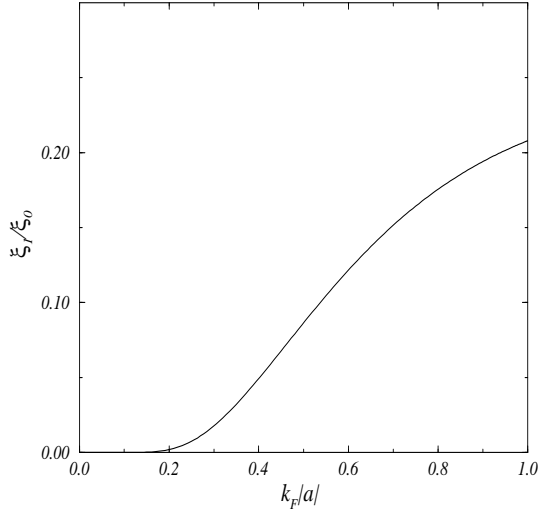,height=8cm,width=8cm}}
}
\caption{The ratio $\xi_1/\xi_0$ of the Kramer-Pesch estimate for 
the vortex core size to the BCS coherence length. }
\label{fig:fig1}
\end{center}
\end{figure}

The microscopic properties of vortices in fermion superfluids can be derived from 
the Bogoliubov-de Gennes (BdG) equations \cite{degennes}, which we write as  
\begin{equation} 
{\cal H}_{\rm BdG} \Psi_i({\bf r}) = E_i \Psi_i({\bf r}), 
\label{eq:bdgeq}
\end{equation} 
where 
\begin{equation} 
{\cal H}_{\rm BdG} = \left( \begin{array}{cc} H_o({\bf r}) & \Delta({\bf r}) \\ 
\Delta^*({\bf r}) & -H_o^*({\bf r}) \end{array} \right), 
\label{eq:bdgham}
\end{equation}
with $H_o = -\hbar^2\nabla ^2 /2m_a -\epsilon_F$, 
and $\Psi_i = (u_i,v_i)^T$, where $u_i$ and $v_i$ are the coefficients 
in the Bogoliubov quasiparticle transformation, and $i$ symbolizes the quantum 
numbers characterizing the quasiparticle states.  
The order parameter $\Delta({\bf r})$ at temperature $T$ is determined by the 
self-consistency condition  
\begin{equation} 
\Delta({\bf r})=\frac{4\pi\hbar^2|a|}{m_a}\sum_i u_i({\bf r}) v^*_i({\bf r})\tanh
\left(\frac{E_i}{2k_BT}\right),  
\label{eq:scpairpot} 
\end{equation} 
where $k_B$ is Boltzmann's constant.  
In cylindrical coordinates $(\rho,\varphi,z)$, a vortex with one quantum 
of circulation can be described by a complex order parameter   
$\Delta({\bf r}) = |\Delta({\bf r})| e^{i\theta({\bf r})}$.  
For a straight vortex line 
along the axis of the container,  
$|\Delta({\bf r})|=\Delta(\rho)$, and 
$\theta = - \varphi$, where $\Delta(\rho)$ starts out at zero in the center of 
the vortex core at $\rho = 0$, and increases asymptotically to the value in 
homogeneous matter $\Delta_0$.
The low-lying quasiparticle 
states can be obtained  analytically, as first 
shown by Caroli {\it et al.} \cite{caroli}.  Kramer and Pesch 
\cite{kramerpesch} used a similar approach in a study of vortices in type II 
superconductors, where they showed that at low temperatures, a new length scale,  
different from $\xi_0$, characterizes the size of the vortex core.   
Their calculation, translated to our notation, is as follows: 
We label the lowest eigenstates of the BdG Hamiltonian by their 
angular momentum $\mu$, equal to half an odd integer, 
and the $z$ component of the 
momentum $k_z = k_F \sin\theta$, where $\theta$ is the angle between 
${\bf k}_F$ and the $xy$ plane.  For $0 < \mu \ll k_F\xi_0$, where 
$k_F\xi_0$ is of order 10-100 in the weak coupling regime, and for 
small $\rho$, one finds that these eigenstates are of the form 
\cite{kramerpesch,degennes,caroli}
\begin{eqnarray}
u_\mu({\bf r}) &=& \left(\frac{k_F\Delta_0}{2L_zv_F}\right)
e^{-i(\mu + 1/2)\varphi}J_{\mu+1/2}(k_F \rho \cos \theta) \label{eq:usol} \\
v_\mu({\bf r}) &=& \left(\frac{k_F\Delta_0}{2L_zv_F}\right)
e^{-i(\mu-1/2)\varphi}J_{\mu-1/2}(k_F \rho \cos \theta) \label{eq:vsol} 
\end{eqnarray} 
where $J_p$ is a cylindrical Bessel function and 
$L_z$ is the length of the vortex line.  We define the characteristic 
length scale of the vortex core via the behavior of the 
order parameter near $\rho = 0$: 
\begin{equation}
\lim _{\rho\rightarrow 0} \Delta({\bf r}) = e^{-i\varphi}\Delta_0 \frac{\rho}{\xi_1}. 
\label{eq:deflength}
\end{equation} 
As $T\rightarrow 0$, the sum in Eq. (\ref{eq:scpairpot}) 
is dominated by the lowest positive energy state, 
$\mu= 1/2$.  Inserting Eqs. (\ref{eq:usol}) and (\ref{eq:vsol}) for $\mu = 1/2$, 
and using $J_p(x) \approx \frac{1}{p!}\left(\frac{x}{2}\right)^p$ for 
$x \ll 1$, we obtain the behavior of the order parameter near $\rho=0$ from 
Eq. (\ref{eq:scpairpot}) as 
\begin{equation}
\Delta({\bf r}) = \frac{\pi|a|}{m_a}\frac{k_F}{L_z v_F}\Delta_0 e^{-i\varphi}\rho  
\sum_{k_z} \cos \theta \tanh \left(\frac{E_{1/2,k_z}}{2k_BT}\right).
\label{eq:step1}
\end{equation}
By comparing with Eq. (\ref{eq:deflength}) we find 
\begin{equation}
\frac{1}{\xi _1} = \frac{\pi k_F|a|}{L_z}\sum_{k_z}\cos\theta 
\tanh\left(\frac{E_{1/2,k_z}}{2k_BT}\right),
\label{eq:step2}
\end{equation}
where we have used $v_F = k_F/m_a$.  Converting the sum over $k_z$ to 
an integral over $\theta$ through the relation 
\begin{equation}
\sum_{k_z} = \frac{L_z}{2\pi}\int dk_z = \frac{L_zk_F}{2\pi}\int_0^\pi d\theta 
\cos\theta,
\label{eq:kzsum}
\end{equation}
we find 
\begin{equation}
\frac{1}{\xi _1} = \frac{k_F^2|a|}{2} \int_0^\pi \cos ^2 \theta d\theta 
= \frac{\pi k_F ^2 |a|}{4}, 
\label{eq:step3}
\end{equation}
since $\tanh (E_{1/2,k_z}/2k_BT) \approx 1$ as $T\rightarrow 0$.  
The characteristic length scale at low temperatures $T\ll T_c$ is therefore 
given by 
\begin{equation}
\xi_1 = \frac{4}{\pi k_F^2 |a|},
\label{eq:charlength1}
\end{equation}
and is seen to be a result of the dominating role played by the low-lying  
eigenstates of the BdG Hamiltonian at low temperatures.    
Numerical  results obtained by Gygi and 
Schl\"{u}ter \cite{gygi2,gygi} have supported this conclusion.    
Furthermore, De Blasio and Elgar\o y \cite{deblasio1,deblasio2}, 
in a study of vortices in 
superfluid neutron matter at $T=0$, considered various ways of defining the 
vortex core size.  They all gave characteristic sizes generally smaller than 
$\xi_0$, the differences being particularly pronounced for low values of 
$\Delta_0$, which is the case in the dilute gas limit.  
The most important feature, however, was that they had a much weaker 
dependence on $k_F|a|$ than the BCS coherence length, and that the overall 
trend of the numerical results was well approximated by 
the Kramer-Pesch estimate for the vortex core size, $\xi _1$. 
The ratio of this quantity to $\xi_0$,  $\xi_1/\xi_0 \approx e^{-\pi/2k_F|a|}/k_F|a|$, 
is plotted in Fig. \ref{fig:fig1}, and can be seen to be much smaller than one  
in the weak coupling regime.  

To estimate the vortex energy per unit length at zero temperature, we follow  
the approach of  \cite{bruun}, but with the important 
modification that we take the vortex core size to be given by 
$\xi_1$ instead of $\xi_0$.  The vortex is modeled 
as a cylindrical region of radius $\xi_1$ containing normal matter, surrounded 
by a velocity field with magnitude given by $v_s = \hbar / 2m_a \rho$.
The superfluid is 
confined within a cylinder of radius $R_c \gg \xi_1$, and is assumed to be uniform 
along the $z$-direction, which also defines the vortex axis.  
The energy per unit length ${\cal E}_v$ of the 
vortex state with respect to the homogeneous superfluid can then be 
divided into two contributions.  The first is the kinetic energy associated 
with the flow of the superfluid outside the vortex core, given by 
\begin{eqnarray}
{\cal E}_{\rm kin} & = & \int_{\xi_1}^{R_c} m_a n_\sigma 
\left(\frac{\hbar}{2m_a \rho}\right)^2 2\pi \rho d\rho \nonumber \\ 
&=& \frac{\pi \hbar^2n_\sigma}{2m_a}\ln\left(\frac{R_c}{\xi_1}\right), \label{eq:emod1}
\end{eqnarray}
and the second contribution, coming from the loss of condensation energy in the 
vortex core, can be estimated 
by multiplying the BCS result for the condensation energy per volume, 
$\epsilon_{\rm cond} = 3\Delta_0^2n_\sigma/4\epsilon_F$, with the 
area of the vortex core $\pi \xi_1 ^2$.  
This gives 
\begin{equation}
{\cal E}_{\rm cond} = \frac{\pi \hbar ^2n_\sigma}{2m_a} 
\frac{3}{\pi^2}\left(\frac{\xi_1}
{\xi_0}\right)^2.
\label{eq:econdfinal}
\end{equation}
The total energy can therefore be written in the usual form 
\begin{equation}
{\cal E}_v = \frac{\pi \hbar ^2 n_\sigma}{2m_a}\ln\left(D\frac{R_c}{\xi_0}\right),
\label{eq:evortex}
\end{equation}
however, in contrast with Eq. (\ref{eq:simplee}), $D$ is now a function of 
$k_F|a|$ through the ratio $\xi_1/\xi_0$.  Since $\xi_1\ll \xi_0$ in the weak 
coupling regime, we have 
\begin{equation}
D = \frac{\xi_0}{\xi_1}\left[1+\frac{3}{\pi^2}\left(\frac{\xi_1}{\xi_0}\right)^2\right],
\label{eq:Dexp}
\end{equation}
and the vortex energy is to a good approximation given by $\ln(R_c/\xi_1)$, 
which has also been found to be the case in neutron matter \cite{deblasio2}.  
Note that $D$ is essentially the inverse of the quantity plotted in Fig. \ref{fig:fig1}, 
and so it increases rapidly with decreasing $k_F|a|$. 
At the limit of the region where weak coupling can reasonably be expected to 
hold, $k_F|a|\sim 0.4$, $D\approx 20.0$, which is an order of magnitude 
larger than in the estimate in Eq. (\ref{eq:simplee}).  

\begin{figure} 
\begin{center}
{\centering
\mbox
{\psfig{figure=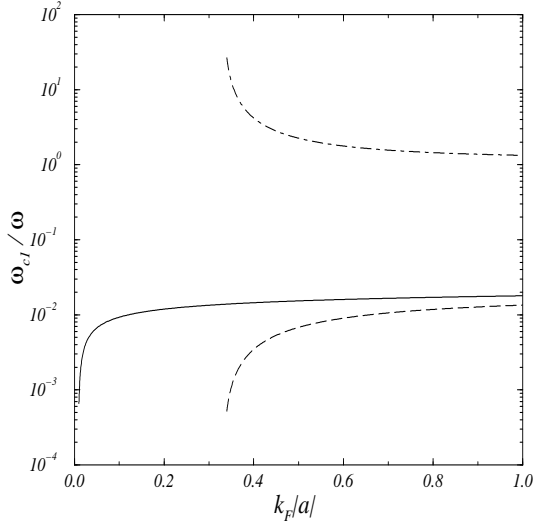,height=8cm,width=8cm}}
}
\caption{The critical frequency, Eq. (\ref{eq:critfreq}) (solid line), the 
critical frequency found in Ref. [10] (dashed line), and their ratio (dot-dashed line), 
for an isotropic trap with the parameters given in the text. }
\label{fig:fig2}
\end{center}
\end{figure}

We can now estimate the critical rotation frequency for   
a dilute gas of atoms trapped in a cylindrically symmetric 
harmonic oscillator potential 
\begin{equation}
V({\bf r}) = \frac{1}{2}m_a\omega_z^2[z^2 + \lambda_T^2(x^2+y^2)],  
\label{eq:trappot}
\end{equation}
where $\lambda_T$ describes the anisotropy of the trap. 
We will use the Thomas-Fermi result for the density profile of this gas, 
given by 
\begin{equation}
n_\sigma(\rho,z)=n_{\sigma,0}\left(1-\frac{\lambda_T^2\rho^2+z^2}{R_z^2}\right)^{3/2},
\label{eq:tfdens}
\end{equation}
where $n_{\sigma,0}=n_\sigma(0,0)$ is the central density of the cloud, 
$R_z= (48N_\sigma\lambda_T^2)^{1/6}l_{\rm osc}$ is the extent of the cloud in the 
$z$-direction, $N_\sigma$ is the number of atoms in the hyperfine state $\sigma$, 
and $l_{\rm osc} = \sqrt{\hbar/m_a\omega_z}$ \cite{butts}. 
Following Lundh {\it et al.} \cite{lundh}, we divide the cloud into vertical 
slices of height $dz$, and use (\ref{eq:evortex}) for a cylinder of radius 
$\rho_1$ such that $\xi_1 \ll \rho_1 \ll R_\perp = R_z/\lambda_T$, where 
the gas can be considered uniform.  The energy per unit length of the slice 
at $z$ can then be written as 
\begin{eqnarray}
{\cal E}_v &=& \frac{\pi\hbar^2 n_\sigma(0,z)}{2m_a}\ln\left(\frac{\rho_1}{\xi_1(z)}
\right) \nonumber \\ 
&+&  \int_{\rho_1}^{R_\perp(z)}m_a n_\sigma(\rho,z)\left(\frac{\hbar}{2m_a\rho}\right)
^2 2\pi \rho d\rho, \label{eq:eslice1}
\end{eqnarray}
with $R_\perp(z) = (1-z^2/R_z^2)^{1/2}R_z/\lambda_T$ being the extent of the 
cloud in the $\rho$-direction for a given $z$, and $n_\sigma(0,z)$ is the 
density along the $z$-axis. Inserting Eq. (\ref{eq:tfdens}) and integrating, 
one finds 
\begin{eqnarray}
{\cal E}_v(z) &=& \frac{\pi\hbar^2 n_{\sigma,0}}{2m_a}\left(1-\frac{z^2}{R_z^2}\right)
^{3/2} \nonumber \\ 
&\times& \ln\left[\frac{2}{e^{4/3}}\frac{R_z}{\lambda_T\xi_1(0)}\left(1-\frac{z^2}
{R_z^2}\right)^{3/2}\right], 
\label{eq:eslice2}
\end{eqnarray}
where $\xi_1(0)$ is the value of $\xi_1$ at the center of the cloud.  
Integrating over $z$, we obtain the total energy of the vortex as 
\begin{equation}
E_v = \frac{\pi \hbar ^2 n_{\sigma,0}}{2m_a}\frac{3\pi}{8}R_z \ln 
\left(0.379\frac{R_z}{\lambda_T\xi_1(0)}\right).
\label{eq:etotal}
\end{equation}
The critical rotation frequency $\omega_{c1}$ for creating a vortex in the trap 
is given by $\omega_{c1}=E_v/L_v$ where $L_v = N_\sigma \hbar$ is the angular 
momentum of the vortex state.  Using $N_\sigma = \pi^2 R_z^3n_{\sigma,0}/
8\lambda_T^2$, we find 
\begin{equation}
\omega_{c1}=\frac{3}{2}\omega_\perp\frac{l_\perp ^2}{R_\perp ^2}
\ln\left(0.379\frac{R_\perp}{\xi_1(0)}\right), 
\label{eq:critfreq}
\end{equation}
where $l_\perp = l_{\rm osc} / \lambda_T$. 
To compare with the result of Ref. \cite{bruun}, we take $\lambda_T=1$ 
so that $\omega_z=\omega_\perp\equiv \omega$, and set $k_{F,0}|a|=0.4$, 
$\epsilon_F=200 \hbar\omega$, 
corresponding to an isotropic trap with $N_\sigma \sim 1.3 \cdot 10^6$.  
With these values, we obtain $\omega_{c1}\approx 0.014\omega$, four times 
larger than in \cite{bruun}.  In Fig. (\ref{fig:fig2}) the ratio between the 
critical frequency found in this paper and the one found in \cite{bruun} is 
plotted as a function of $k_F|a|$, with the other parameters kept at the same 
values as above.  Note that $\omega_{c1}$ turns negative for small values 
of $k_F|a|$, reflecting that the condition $\xi_1 \ll R_\perp$ is violated.  
For the parameters used in this example, this occurs at $k_F|a| \approx 0.01$ for 
$\omega_{c1}$ given by Eq. (\ref{eq:critfreq}), and for $k_F|a| \approx 0.33$ for 
the corresponding result in Ref. \cite{bruun}.  

To conclude, we have estimated the energy of a vortex in a uniform, dilute 
gas of atoms at $T=0$.  The quasiparticle states in the core of the vortex introduce 
a new characteristic length scale $\xi_1$ which has a weaker density dependence 
than the BCS coherence length $\xi_0$ and is much smaller than $\xi_0$ in 
the dilute gas regime $k_F|a| \ll 1$.  Taking this fact into account leads us 
to predict a considerably higher vortex energy than what was found in earlier 
calculations.  For the case of a trapped gas of fermionic atoms, this means that 
the critical frequency for the formation of a vortex state can be $5 - 10$  
times higher than previously estimated.

The author is grateful to   F. V. de Blasio, G. M. Bruun, R. Kaldare, 
and C. J. Pethick for illuminating and valuable discussions.  
Support through a post-doctoral fellowship from The Research Council of 
Norway (NFR) is acknowledged.

\end{multicols}

\end{document}